\documentstyle[aps,psfig]{revtex}
   \newcommand{\"}{\"{o}}
   \newcommand{\"}{\"{a}}

   \newcommand{\cent}[1] {\begin{center}#1\end{center}}
   \newcommand{\beq} {\begin{equation}}
   \newcommand{\eeq} {\end{equation}}
   \newcommand{\bdm} {\begin{displaymath}}
   \newcommand{\edm} {\end{displaymath}}
   \newcommand{\beqa}{\begin{eqnarray}}
   \newcommand{\eeqa}{\end{eqnarray}}
   \newcommand{\bean}{\begin{eqnarray*}}
   \newcommand{\eean}{\end{eqnarray*}}
   \newcommand{\bea}{\begin{array}}
   \newcommand{\eea}{\end{array}}
   \newcommand{\betab}{\begin{tabbing}}
   \newcommand{\eetab}{\end{tabbing}}
   \newcommand{\bma}{\begin{math}}
   \newcommand{\ema}{\end{math}}
   
   \newcommand{\vecbm}[1]{\mbox{\boldmath#1}}

   \newcommand{\mra}  {\rightarrow}

\begin{document}
\twocolumn
\draft
%\tighten
\title{
Microcanonical vs. canonical thermodynamics }
\author{
D.H.E. Gross and M.E. Madjet}
\address{
Hahn-Meitner-Institut
Berlin, Bereich Theoretische Physik,Glienickerstr.100\\ 14109 Berlin, Germany
and 
Freie Universit\"at Berlin, Fachbereich Physik}
\maketitle
\begin{abstract}
The microcanonical ensemble is in important physical situations different from
the canonical one even in the thermodynamic limit.  In contrast to the
canonical ensemble it does not suppress spatially inhomogeneous configurations
like phase separations. It is shown how phase transitions of first order can be
defined and classified unambiguously for finite systems without the use of the
thermodynamic limit. It is further shown that in the case of the 10-states
Potts model as well for the liquid-gas transition in Na, K, and Fe the
microcanonical transition temperature, latent heat and interphase surface
tension are similar to their bulk values for $\sim 200-3000$ particles. For Na
and K the number of surface atoms keeps approximately constant over most of the
transition energies because the evaporation of monomers is compensated by an
increasing number of fragments with $\ge 2$ atoms (multifragmentation).
\end{abstract}
In the thermodynamic $\lim_{N\mra\infty}|_{N/V=\varrho}$
the microcanonical ({\em ME}) and the (grand)canonical ensembles ({\em CE}) are
usually considered to be identical. As this is not the case e.g. at phase
transitions of first order and as both ensembles describe different physical
situations in the case of finite systems it is necessary to emphasize here the
differences between the two. It is  important to realize that in the
microcanonical ensemble phase transitions can very well be defined and
classified for finite systems (sometimes for systems of some hundred particles)
without the use of the thermodynamic limit. It is thus possible to define phase
transitions even in systems which have no thermodynamic limit at all like
systems with unscreened forces of long range.  The arguments are not entirely
new, however, many discussions with workers from different physical disciplines
showed that it is necessary to state these facts clearly.
 
The difference between the {\em ME} and the {\em CE} can be seen most easily
for the probability $P(E/N)$ in the $\em CE$ at a phase transition of first
order ($T=T_{tr}$) e.g. liquid to gas. Here $P(E/N)$ is a bimodal distribution
with a peak at the specific energy of the liquid $E_l/N$ and another at the
specific energy of the gas $E_g/N$. As $E_g/N-E_l/N=q_{lat}$ the specific
latent heat, the fluctuations of the total energy {\em per particle} remain
finite even in the thermodynamic limit. Consequently, the canonical is
different from the microcanonical ensemble.

The {\em ME} is the ensemble which is obtained directly from the mechanics of
the N-body system. The classical partition sum $W(E)$ is the volume
of the energy shell in the N-body phase space in units of $(2\pi\hbar)^{3N}$.
In the quantum case it is the number of N-body states at the energy $E$. The
canonical partition sum is obtained from $W(E=N\varepsilon)$ by the Laplace
transform
\begin{equation} 
Z(T,P,N)=\int \int_0^{\infty}{W(E,V,N)e^{-N(\varepsilon+Pv)/T}\;dE\;dV}.
\label{laplace}
\end{equation}
 
As the Laplace transform is a very stable transformation, the inverse
is highly unstable. Uncertainties or ``reasonable'' approximations of the
canonical $Z(T)$ can imply serious defects in the microcanonical $W(E)$. An
example was discussed for the Bethe nuclear level-density formula in
\cite{gross124}. In this respect the {\em ME} is the fundamental ensemble.  The
epigraph on Boltzmann's gravestone is the most concise formulation of
thermodynamics: \cent{\fbox{\fbox{\vecbm{$S=k*lnW$}}}}

But there is another reason why the {\em ME} applies for  much more physical
situations than the {\em CE}. The microcanonical caloric function $T(E/N)$
depends usually only very weakly on the number of particles $N$ and reflects
already for a couple of some $100$ particles quite often the bulk properties.
Moreover, the {\em ME} can be applied to {\em inhomogeneous} systems
like e.g. when there are forces with a range longer than the linear dimensions
of the system. This is the situation in selfgraviting systems but also small
systems like hot nuclei are temporarily equilibrated systems under the long
range Coulomb force.  Statistical multifragmentation of hot nuclei must be
and was treated by microcanonical statistics from its early days, see
\cite{gross95,gross155}. The {\em CE} has the tendency to {\em suppress
inhomogeneities the larger the number of particles}\cite{gross153}, see below.
It is interesting to notice that Georgii has shown that if one enforces
translational invariance of the {\em ME}, i.e.  if one projects out
inhomogeneous configurations (with phase separations)
\cite{georgii95} it becomes equivalent to the {\em CE} in the thermodynamic
limit also at phase transitions of first order.

At phase transitions of first order the system becomes inhomogeneous
and fragments into spatially separated pieces of ``liquid'' and
``gas'' phases. To see how the {\em ME} handles this situation we discuss the
Potts model \cite{potts52,binder76}, a generalization of the Ising
model, defined by the Hamiltonian:
\begin{equation}
H=\sum_{i<j}{'\{1-\delta_{\sigma_i,\sigma_j}\}}
\end{equation}
on a two dimensional lattice of
$N=L*L$ spins here with $q=10$ possible components.  The sum is over pairs of
nearest neighbor lattice points only. The Potts model can easily be simulated
numerically and its behavior in the thermodynamic limit is known analytically
\cite{baxter73}.  The microcanonical partition sum  over all possible different
configurations $\nu$ with the same total energy $E$ is:
\begin{equation}
W(E)=\sum_{\nu}{\delta_{E_{\nu},E}} \;\;.
\end{equation}

\begin{figure}
\centerline{\psfig{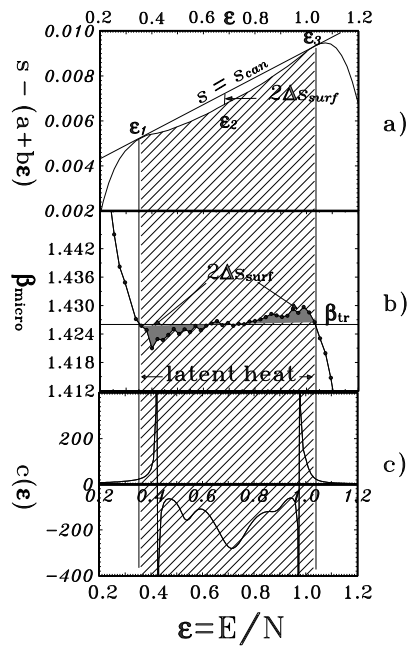}}
%\includegraphics*[bb = 36 30 415 622, angle=-180, width=7.5cm,  %<-- altern. method
%clip=true]{cris2np.eps}                                         %<--
%% the four numbers are x-lower-left y-lower-left x-upper-right y-upper-right
%%    in postscript units as read from ghostview
\caption{Microcanonical phase transition of first order in the Potts model,
$L=200$, $q=10$.}
\label{potts} 
\end{figure}

Fig.\ref{potts}b shows the microcanonical caloric equation
$1/T=\beta(\varepsilon=E/N)$, obtained by simulating $\beta=\partial s/\partial
\varepsilon$ in $\approx12*10^{5}$
microcanonical sweeps per energy step for a $L=200$ lattice. The phase
transition is manifested by the backbending of $\beta(\varepsilon)$
\cite{gross150}. The ``Maxwell'' line defines the transition temperature and
the two shaded areas below(above) are just the entropy loss (gain) $\Delta
s_{surf}$ when one separates(joins) the two phases by an interphase surface.
Spins on the surface have a reduced freedom and consequently the entropy of the
lattice is reduced per surface spin by $\Delta s_{surf}*L=\sigma_{surf}/T_{tr}$
the specific surface tension. The length of the ``Maxwell'' line
$\varepsilon_3-\varepsilon_1$ is the latent heat $q_{lat}$ per particle.  This
S-bending of the microcanonical caloric curve $\beta(\varepsilon)$ is a signal
of a phase transition of first order \cite{gross153,gross154,gross155}.   

The total entropy per lattice point is given by
$s(\varepsilon)=\int_0^{\varepsilon}{\beta(\varepsilon') d\varepsilon}'$
(fig.\ref{potts}a). In order to visualize the anomaly of the entropy a linear
function $a+b\varepsilon$ ($a=0.17$, $b=1.42$) was subtracted. The depth of the
convex intruder is again the loss of entropy $\Delta s_{surf}$ when a surface
is created to separate the two phases. As we use periodic boundary conditions
one needs two cuts to separate the phases. 

The specific heat $c(\varepsilon)=\frac{\partial \varepsilon}{\partial<T>}
=-\beta^2\frac{\partial\beta}{\partial\varepsilon}$ 
is shown in fig.\ref{potts}c after smoothing the statistical fluctuations of
$\beta(\varepsilon)$. $c(\varepsilon)$ becomes {\em negative} in the 
shaded region whereas $c(T)=N(<\varepsilon^2>-<\varepsilon>^2)/T^2$ must be
positive in the {\em CE}.  The canonical ensemble of the bulk jumps over the
shaded region between the vertical lines at $\varepsilon_1$ and
$\varepsilon_3$. This is the region of coexistence of two phases one with
ordered spins, the other with disordered spins. Here $c(\varepsilon)$ has two
poles and becomes negative in between.  Notice that the poles are {\em inside}
$\varepsilon_1\le\varepsilon\le\varepsilon_3$, i.e. the canonical specific heat
remains finite and positive as it should.

In the {\em CE} configurations with different energies contribute 
$\propto exp[-N*f(\varepsilon)/T]$ to the canonical partition sum with
$f(\varepsilon)=\varepsilon-Ts(\varepsilon)$ the free energy per particle.  
Due to the convex intruder in $s(\varepsilon)$ by $\Delta s_{surf}$
{\em configurations with coexistent phases and energies $\varepsilon\sim
\varepsilon_2$ are suppressed in the {\em CE} by the exponential factor
$exp(-N\Delta s_{surf})=exp(-\sigma_{surf}L/T)$.} This fact is frequently used
in canonical model simulations of phase transitions to obtain the surface
tension e.g. \cite{binder82,janke94}.

\begin{figure}
\centerline{\psfig{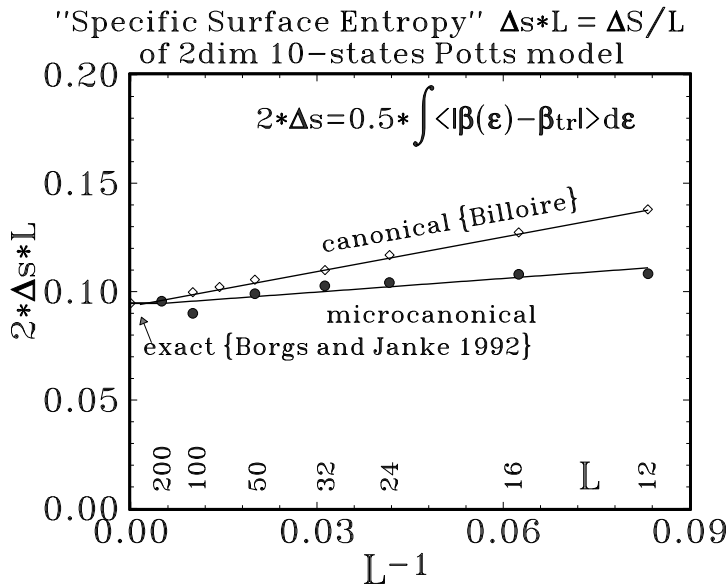}}
%\includegraphics*[bb = 2 10 491 613, angle=-90, width=7cm,  %<-- altern. method
%clip=true]{pr32p.eps}\\                                         %<--
%% the four numbers are x-lower-left y-lower-left x-upper-right y-upper-right
%%    in postscript units as read from ghostview
\caption{Surface tension as function of $1/L$. 
\protect\cite{gross150,borgs92,billoire93}}
\label{pr32tamc} 
\end{figure}

In Fig.\ref{pr32tamc} the microcanonical surface tension is shown as function
of $1/L$. The surface tension deviates for a lattice of $L=12$ by only $10\%$
from the bulk values. Other parameters deviate even less from the bulk value,
e.g. the transition temperature by less than $2\%$. {\em Within the {\em ME} a
phase transition can be identified with all its salient parameters, the
transition temperature, the specific latent heat, 
the interphase surface
tension, and the change of entropy per particle. It can unambiguously be
identified and classified for finite systems from its microcanonical caloric
equation of state $T(E/N)$. There is no need to refer to the thermodynamic
limit.  Evidently, the usual signal for a phase transition as a singularity in
a canonical observable as function of the temperature is artificial and applies
to the canonical ensemble only, c.f. e.g.\cite{hueller94,gross153,gross155}.} This
conclusion has far reaching consequences: One can easily extend the concept of
a phase transition to systems which do not have any thermodynamic limit like
e.g.  selfgraviting systems or charged systems under unscreened Coulomb
repulsion like hot nuclei \cite{gross95}.

\begin{figure}
\centerline{\psfig{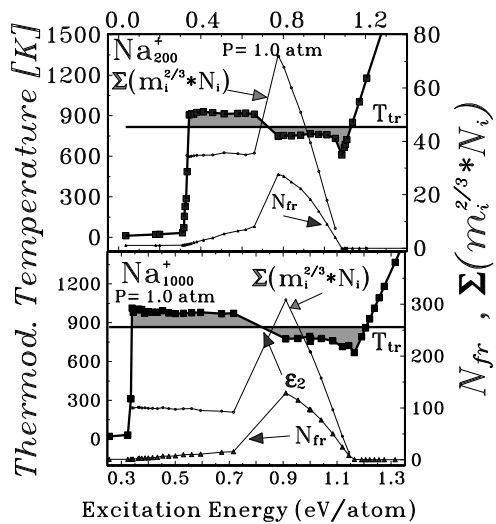}}
%\includegraphics*[bb = 32 147 484 570, angle=-90, width=7cm,  %<-- altern. method
%clip=true]{NaPRL.eps}                                         %<--
% the four numbers are x-lower-left y-lower-left x-upper-right y-upper-right
%    in postscript units as read from ghostview
%\caption{Caloric curve (thick line), number of fragments
%$N_{fr}$ with $m_i\ge2$, and number of surface atoms $\sum(m_i^{2/3}N_i)$ of
%microcanonical systems with 200 and 1000 sodium atoms at normal pressure.
\caption{Caloric curve (thick), number $N_{fr}$ of fragments with $m_i\ge2$,
and ``number'' of surface atoms $\sum(m_i^{2/3}N_i)$ of microcanonical systems
with 200 and 1000 sodium atoms at normal pressure.}
\label{NaPRL}
\end{figure}

In order to test the relation of the area under the backbend of
$\beta(\varepsilon)$ to the empirical bulk surface tension for a realistic
system, we simulated the phase transition for evaporation and
multifragmentation of some hundreds Na, K, and Fe atoms under normal
pressure by our microcanonical Metropolis Monte Carlo sampling {\em MMMC}. The
phase transition shows up as a clear anomaly (backbend) of the microcanonical
caloric curve $T(\varepsilon)$ c.f. fig.\ref{NaPRL}. It is easy to determine
the transition temperature $T_{tr}$, the latent specific heat $q_{lat}$ and the
surface entropy per particle $\Delta s_{surf}$. The transition is clearly
identified to be of first order. In order to calculate the surface tension one
has to determine the total surface area of all fragments. This is somewhat
difficult as the size of the fragments drops strongly with rising excitation
energy because of an increasing evaporation of monomers. In the fragmentation
of Na, K clusters the total surface area keeps nevertheless roughly constant
and close to $N_0^{2/3}$
in the first part of the transition in $\varepsilon_1\le \varepsilon \le
\varepsilon_2$. Here the reducing size $m_i$ of the fragments is compensated by
an increasing number of fragments $N_{fr}=\sum N_{mi\ge2}$. Therefore, the
surface tension may be estimated by formula (\ref{sig}), where the average is
taken over the energy interval $\varepsilon_1\le \varepsilon\le \varepsilon_2$.
As iron decays by pure monomer evaporation because of its much larger surface
tension, there is only one evaporation residue with a steadily diminishing
surface area and the use of formula (\ref{sig}) to determine the surface
tension is problematic.  The values of the so determined surface tension
approach systematically the bulk values, table \ref{table1}.
\begin{eqnarray}
\sigma_{surf}&\sim& T_{tr}*\Delta s_{surf}*N_0/N_{eff}^{2/3}\nonumber \\
N_{eff}^{2/3}&=&<\!\!\sum_{m_i\ge 2}({m_i^{2/3}*N_i)\!\!>},\label{sig}
\end{eqnarray}

Conclusion: Whether an interacting many-body system has a phase transition is
not a property of an infinitely large system in the thermodynamic limit. One
can see and classify phase transitions in finite systems by the form of the
microcanonical caloric equation of state $T(E/N)$.  This opens the possibility
to define and discuss phase transitions also in systems which are not
``thermodynamically stable'' in the sense of van Hove \cite{vanhove49}.  {\em
MMMC} \cite{gross141,gross153} allows to calculate $T(E/N)$ for some $100$ to
$1000$ particles. For our examples the transition represents quite well the
properties of the bulk. Configurations with coexistent, separated phases are
well represented in the{\em ME} whereas in the {\em CE} they are suppressed by
a factor $exp(-N^{2/3}\sigma_{surf}/T_{tr})$.
\begin{table}[htb]
\begin{tabular}{|c|c|c|c|c||c|} \hline
&$N_0$&$200$&$1000$&$3000$&bulk\\
\tableline 
\hline 
&$T_{tr} \;[K]$&$816$&$866$&&$1156$\\ \cline{2-6}
&$q_{lat} \;[eV]$&$0.791$&$0.871$&&$0.923$\\ \cline{2-6}
{\bf Na}&$s_{boil}$&$11.25$&$11.67$&&$9.267$\\ \cline{2-6}
&$\Delta s_{surf}$&$0.55$&$0.56$&&\\ \cline{2-6}
&$N_{eff}^{2/3}$&$39.94$&$98.53$&&$\infty$\\ \cline{2-6}
&$\sigma/T_{tr}$&$2.75$&$5.68$&&$6.34$\\
\hline
\hline
&$T_{tr} \;[K]$&$697$&$767$&&$1033$\\ \cline{2-6}
&$q_{lat} \;[eV]$&$0.62$&$0.7$&&$0.80$\\ \cline{2-6}
{\bf K}&$s_{boil}$&$10.35$&$10.59$&&$8.99$\\ \cline{2-6}
&$\Delta s_{surf}$&$0.65$&$0.65$&&\\ \cline{2-6}
&$N_{eff}^{2/3}$&$32.52$&$92.01$&&$\infty$\\ \cline{2-6}
&$\sigma/T_{tr}$&$3.99$&$7.06$&&$7.40$\\
\hline
\hline
&$T_{tr} \;[K]$&$2600$&$2910$&$2971$&$3158$\\ \cline{2-6}
&$q_{lat} \;[eV]$&$2.77$&$3.18$&$3.34$&$3.55$\\ \cline{2-6}
{\bf Fe}&$s_{boil}$&$12.38$&$12.68$&$13.1$&$13.04$\\ \cline{2-6}
&$\Delta s_{surf}$&$0.75$&$0.58$&$0.77$&\\ \cline{2-6}
&$N_{eff}^{2/3}$&$28.34$&$61.41$&$156.94$&$\infty$\\ \cline{2-6}
&$\sigma/T_{tr}$&$5.29$&$9.44$&14.72&$16.57$\\
\hline
\end{tabular} 
\caption{Parameters of the liquid -- gas transition at $1$ atm. in a
microcanonical system of $N_0$ interacting atoms and in the bulk. The bulk
values of $\sigma$ are calculated by $\sigma=4\pi r_{WS}^2(a_s-b_sT_{tr})$ with
$a_s$ consistent to the value used in this calculation for the binding energies
of the cluster. It was determined by \protect\cite{brechignac95b} from
experimental binding energies of different clusters. It is larger than the bulk
values of \protect\cite{miedema78}. As $b_s$ for clusters is not known and not
used in this calculation it was taken from \protect\cite{miedema78}. For
multifragmenting Na and K the calculations are considerably more time consuming
prohibiting the calculation for $N_0=3000$.\label{table1}}
\end{table}

%\cite{brechignac95b}
%\cite{miedema78}
%\bibliographystyle{unsrt}%{alpha}%{plain} %{unsrt}
%\bibliography{c:/bibliogr/gross,c:/bibliogr/othbiba,c:/bibliogr/othbibb,c:/bibliogr/othbibcd,c:/bibliogr/othbibe,c:/bibliogr/othbibf,c:/bibliogr/othbibg,c:/bibliogr/othbibh,c:/bibliogr/othbibij,c:/bibliogr/othbibk,c:/bibliogr/othbibl,c:/bibliogr/othbibm,c:/bibliogr/othbibn,c:/bibliogr/othbibo,c:/bibliogr/othbibp,c:/bibliogr/othbibr,c:/bibliogr/othbibs,c:/bibliogr/othbibt,c:/bibliogr/othbibuw,c:/bibliogr/othbibxz}
%\end{document}

\end{document}